# Increasing System Test Coverage in Production Automation Systems

Sebastian Ulewicz*, Birgit Vogel-Heuser

Technical University Munich, Institute of Automation and Information Systems, Boltzmannstr. 15, 85748 Garching near Munich, Germany

* Corresponding author; email: sebastian.ulewicz@tum.de, fax: +49 89 289 16410

## Abstract

An approach is introduced, which supports a testing technician in the identification of possibly untested behavior of control software of fully integrated automated production systems (aPS). Based on an approach for guided semi-automatic system testing, execution traces are recorded during testing, allowing a subsequent coverage assessment. As the behavior of an aPS is highly dependent on the software, omitted system behavior can be identified and assessed for criticality. Through close cooperation with industry, this approach represents the first coverage assessment approach for system testing in production automation to be applied on real industrial objects and evaluated by industrial experts.

## 1 Introduction

Automated production systems (aPS) in factory automation have high requirements regarding availability and reliability (Vogel-Heuser, Fay, Schaefer, & Tichy, 2015), as these systems typically run over long periods of time (decades) and system failures or incorrect behavior can increase costs. The volume and complexity of aPS's software has risen substantially over the last decade (Vyatkin, 2013), exacerbating the problem of ensuring reasonable system quality. This quality is typically investigated and assured by testing. Apart from unit tests performed on single software modules in an early design phase, system tests of the integrated functionality of software and hardware are defined and performed in late phases of development, often as late as during on-site plant commissioning.

From the authors' experience, test plans for system testing exist in most companies in the field of automated production systems engineering, yet the definition of the individual test cases is abstract and generic. On the one hand, large parts of these test plans can be reused between projects, on the other hand, the individual test cases leave a lot of room for interpretation during the testing process. Additionally, tests are performed manually, as many functions are not related to the software alone but also to the integrated system comprised of mechanical and electrical hardware as well as software. Thus, many actions performed during these tests, such as placing intermediate products into the machine and visually verifying the correct product quality, cannot be performed fully automatically: Sensors and actuators that would enable automated testing are not available due to their cost. Instead, the test operator is required to perform these actions manually. This testing process is often performed under high time pressure in an uncomfortable on-site environment and based on the mentioned vaguely specified requirements on the system. This results in uncertainty of the adequacy of the performed tests: The adequacy of the performed tests to ensure the abstractly defined required functionality is often based on the experience and intuition of the test operator. Subsequently, the possibility of not testing critical behavior and thus overlooking critical faults in the system represents a realistic problem.

Code coverage is a possibility to assess test adequacy (Zhu, Hall, & May, 1997). As the behavior of the integrated automated system is largely dependent on the software, a coverage assessment of implemented behavior can be performed: by identifying uncovered (untested) code, unintended omissions of testing system behavior can be revealed. Based on this finding, an approach was developed consisting of an instrumentation of the control software to allow recording of execution traces and an analysis of these traces for coverage assessment and identification of untested code. The approach was implemented in a prototypical tool and evaluated using a real industrial case study and a subsequent expert evaluation yielding promising results.

The main contribution of the presented approach is the ability to identify untested behavior during system testing of fully-integrated industrial production automation systems controlling discrete processes without the need for formalized requirements or simulations. Thus, for the first time, the approach provides valuable support in quantitatively assessing and increasing testing quality in fully-integrated industrial aPS in industrial quality assurance scenarios.

The structure of the paper is as follows: In section 2, an overview of requirements gathered from industrial experts is presented in order to rate existing approaches and to guide the development of the presented approach. Section 3 presents related work in the field of production automation and adjacent domains. These works are analyzed and a research gap is identified. In the concept section (section 4), the approach is described in detail regarding the choice of the coverage metric, the code instrumentation and coverage assessment. Information about the implementation of the approach is detailed in section 5 followed by an application on a case study and an expert evaluation in section 6 which allows a qualitative impression of the concept's performance and applicability. In the last sections (7 and 8), a conclusion and an outlook on future work are discussed.

## 2 Industrial Requirements Regarding System Testing in Production Automation

The aim of the approach presented in this work is to be closely aligned with industrial requirements in the field of production automation engineering. For this reason, multiple workshops with up to seven highly experienced experts from three internationally renowned companies active in this field were conducted to infer these requirements. Six main requirements regarding applicability of a possible approach were identified:

### Requirement 1 (R1): Support of Industrial Software Properties
Programming standards relevant for the industry, i.e. the programming standard IEC 61131-3 for Programmable Logic Controllers (PLC) dominant in the engineering of automated production machines, has to be supported. Of the five defined programming languages, Structured Text (ST) and Sequential Function Chart (SFC) in particular need to be supported as these are the most commonly used programming languages within the companies questioned.

### Requirement 2 (R2): Real Time Capability and Memory Size
The approach should not influence the real time properties of the tested system in a way that would not permit needed real time capabilities of the system to hold. The needed real time capabilities are seen as unaffected if a possible increase in execution time of modified code does not lead to the PLC scan cycle time to be exceeded. In addition, possibly increased size of compiled control code software should not lead to exceeded memory on the execution hardware (PLC).

### Requirement 3 (R3): Inclusion of Hardware and Process Behavior
Testing a system integrates all parts of the system, meaning software, hardware and the controlled process. To be able to assess a system's conformance to its specification, all parts should be as similar to the final system as possible, i.e. the software running on the final execution hardware, controlling the final version or the hardware setup and technical process. For this reason, using a simulation rather than the real hardware is often not sufficient for final system tests, as the validity of the described behavior is a simplification of real hardware behavior. In addition, simulations are costly to produce – in particular for aPS produced in small lot sizes – and automatic generation of simulations with available documents as proposed by (Barth & Fay, 2013; Puntel-Schmidt et al., 2014) are not available in many cases and for the participating industry partners in particular. This problem especially applies to medium and smaller sized companies, where an approach which is independent of simulations is required, as these are often no option for system testing in production automation for economic reasons.

Requirement 4 (R4): Manipulation of Hardware and Process Behavior

The approach needs to be applicable on real industrial testing use cases, as defined by the currently performed system test cases in the company. System tests, as described in this approach, are defined as black box tests (test derived from a specification rather than the code itself) of a fully integrated system comprised of software, controlled hardware and the technical process. The tests include manual manipulations of the hardware or technical process that cannot be performed by the software. As an example, manually opening and closing doors or putting intermediate products in the machine can be typical operations during system testing.

Requirement 5 (R5): No Need for Formalized Functional Requirements

The problems stated in the introduction could be mitigated using more detailed and formalized functional requirements. Using a connection between requirements, test cases and models of different engineering views of the system could enable validation of the involved models (Estevez & Marcos, 2012) and more detailed relation between requirements and a system's software code itself could be created using static feature location techniques (Dit, Revelle, Gethers, & Poshyvanyk, 2013). Yet in practice, this would require adequate software tools, substantial effort regarding training and additional resources for specification for each new engineering project. As this tradeoff between an initial investment and its outcome is highly speculative, according to the participating industrial experts, the approach must be independent from formalized functional requirements.

Requirement 6 (R6): Support the Assessment of Test Adequacy – Finding Untested Behavior

Here, the approach is to increase efficiency and quality during the quality assurance process of special purpose machinery by supporting the tester, who might be experienced software engineers or inexperienced technicians, when assessing the test adequacy. A generic coverage assessment, i.e. "100% of behavior has been tested", is seen as questionable because a resource for completely testing a system is not feasible and specific numbers may have little meaning. Therefore, rather than assessing how complete the system behavior was tested, the requirement was set to finding untested behavior and assessing its need for specifying tests.

## 3    Related Work in the Field of Test Coverage Assessment

Coverage metrics in the field of computer science have been an active research topic for many years. They can be used for test case generation (Anand et al., 2013), change impact analysis (Bohner & Arnold, 1996; De Lucia, Fasano, & Oliveto, 2008), regression test selection and prioritization (Engström & Runeson, 2010; Yoo & Harman, 2012) or for assessing test suite adequacy (M. C. K. Yang & Chao, 1995; Zhu et al., 1997). While some approaches have already been incorporated into the production automation domain, coverage metrics have rarely been used for assessing test suite adequacy in this field. In the following, a closer look into work related to the presented approach will be taken.

## 3.1    Requirement Based Test Coverage

These coverage metrics are based on the relation of requirements and test cases in which test cases check whether the system under test fulfils a set of requirements. In reverse, if an approach uses functional requirements or specifications for test generation, it is assumed that the generated test case is adequate for these requirements.

A basic realization of this approach is commercially available in multiple requirements management tools, such as IBM Rational DOORS (IBM, 2016) or Siemens Polarion (Siemens, 2016): informally specified requirements can be linked to informally specified test cases. If a requirement does not have a related test case, it is assumed that a test case is missing. In case one or more test cases are linked to a requirement, the requirement is seen as fulfilled if all test cases were completed successfully. This implies that the creator of the test cases specified all relevant test scenarios, which is relying heavily on the ability of the individual. If test cases for a requirement were missing, this would not become apparent if all other test cases were executed successfully as no quantitative measure beyond the connection of tests and requirements is given.

Using formalized requirements, the expected behavior of the system can be specified in more detail and a subsequent relation between content of the requirement and the test cases can be performed, rather than solely evaluating the connection of requirement and test case. This was implemented in different research works in specific applications for embedded systems (Siegl, Hielscher, German, & Berger, 2011), computer science (Whalen, Rajan, Heimdahl, & Miller, 2006) and testing of safety field busses (Jan Krause, Hintze, Magnus, & Diedrich, 2012). In production automation, multiple approaches in the field of model based testing have generated test cases from formalized specifications (Rösch, Ulewicz, Provost, & Vogel-Heuser, 2015). Some use environmental models of the system (Kumar, Gilani, Niggemann, & Schäfer, 2013) or special requirement ontologies (Sinha, Pang, Martínez, & Vyatkin, 2016) to generate executable test cases which requires these complex models to be specified and validated. Some works use modified UML sequence diagrams to specify test cases (Hametner, Kormann, Vogel-Heuser, Winkler, & Zoitl, 2011; Kormann & Vogel-Heuser, 2011), although this approach provides no information about coverage of the defined test case. More recently, an approach using timing sequence diagrams to generate test cases covering all possible signal mutations for different classes of mutations was proposed (Rösch & Vogel-Heuser, 2017). Although a subsequent selection of relevant test cases can be performed, the resulting coverage is not calculated.

An approach which bases test coverage on formalized or traceable requirements exhibits multiple positive aspects: 1) it is a direct measure of how well a test suite addresses a set of requirements, 2) it is implementation independent and 3) it does not require execution or instrumentation of the system under test (Whalen et al., 2006). At the same time, this type of metric has several negative aspects:

1. Detailed requirements specification as well as a model of the relation between these requirements and test cases need to be available, this requires additional effort.
2. A quantitative evaluation of this metric requires formalized requirements specifications or relies solely on the notion that a given set of test cases can fully strengthen the notion that all requirements are fulfilled, even if each requirement only relates to a single test case.
3. Missing or incomplete requirements will go unnoticed in this metric. If the set of requirements is not maintained correctly or inadequately defined from the beginning, the metric cannot yield satisfying results.
4. Unrequired parts of the code cannot be identified as test cases are related to requirements only. If unneeded code was implemented during code implementation, this metric is unable to identify these unnecessary parts of the code resulting in additional maintenance effort in later stages of system maintenance.

## 3.2 Code Structure Based Test Coverage

Structural code coverage metrics have been a common method for assessing software test adequacy in safety critical systems: For safety critical avionics systems, the DO-178b standard proposes the use of structural metrics for assessing the adequacy of a suite of tests for a test subject (RTCA, 1992). Depending on the criticality, the standard proposes more or less detailed metrics, such as statement coverage ("every statement in the program has been executed at least once" (RTCA, 1992)) or condition/decision coverage ("every point of entry and exit in the program has been invoked at least once, every condition in a decision in the program has taken all possible outcomes at least once, and every decision in the program has taken all possible outcomes at last once" (RTCA, 1992)). While the presented approach does not aim to fulfil coverage criteria as in software testing for safety critical systems for economic reasons, certain properties about the different metrics still apply to the field of aPS. The different metrics differ substantially regarding the number of test cases to fulfill each criterion, but also in their ability to detect faults. In most cases, full statement coverage needs fewer test cases but can fail to detect errors in complex decisions within the control flow. The practical ability to detect faults in desktop software was evaluated for unit testing (Zhu et al., 1997) and complete test suites (Gligoric et al., 2013; Gopinath, Jensen, & Groce, 2014). From these evaluations, even the simple statement coverage metric turns out to be a very valuable metric, especially for finding out if a test suite is inadequate (missing test cases). The DO-178b (RTCA, 1992) also proposes statement coverage as the minimal requirement for safety critical systems.

There are many tools available from computer science for structural based coverage analysis. An overview of available tools is given by (Q. Yang, Li, & Weiss, 2009), yet not all tools have remained in development. Still, tools such as Atlassian Clover (Atlassian, 2016), BullseyeCoverage (Bullseye, 2016) and Unicom PurifyPlus (Unicom, 2016) are readily available and offer coverage analysis using multiple coverage criteria for higher object-oriented programming languages (e.g. Java, C++, C#). Even if their application on the programming languages and execution hardware of PLCs could be achieved, these tools were not developed in respect to the industrial scenarios required in the aPS industry: all tests are executed fully automatic and do not require human operators. In addition, the influence of the tracing algorithms used by the tools is unsure regarding a port into the PLC field.

In comparison to computer science and critical embedded systems, production automation engineering has seen few comparisons of structural code coverage metrics for test adequacy assessment. A tool for test coverage measurement for Function Block Diagrams (FBD) is presented by (Jee, Kim, Cha, & Lee, 2010) for use in safety critical programs of nuclear reactors. The test case coverage is externally checked by analyzing the data flow paths of the FBD and comparing them to the test inputs. This approach seems to be hardly applicable to automated production systems: discrete process step chains are prevalent in automated production systems which differ substantially from the complex logical data flows of nuclear power plants. A reason for the hesitant use of coverage criteria in production automation might be the strict real time requirements and the overhead created by measuring coverage. Only few works present efficient tracing algorithms for embedded systems (Wu, Li, Weiss, & Lee, 2007) and automation software (Berger, Prähofer, Wirth, & Schatz, 2012; Prähofer, Schatz, Wirth, & Mössenböck, 2011). As tracing algorithms are designed for embedded systems, the scenarios and test environment prohibit an easy adaption to automated production systems. The approach for automation software uses a very sophisticated tracing approach aimed at debugging automated production systems, not taking coverage assessment into account. In particular, this approach does not include structured system tests or a connection of tests to the recorded data, and focuses mainly on reproducing variable values at certain points in time for debugging purposes.

Some works in the production automation field take a different approach stemming from computer science: testing input sequence generation from the code itself to achieve full coverage related to a certain criterion (Jee, Yoo, & Cha, 2005; Simon et al., 2015). Other works build on this approach for software product lines offering efficient test input sequence generation techniques (Bürdek et al., 2015; Lochau et al., 2014). These approaches possess two main problems for system testing in production automation: 1) the generation technique generates test input sequences directly from the code, but does not include an expected behavior ("test oracle"). The test cases themselves are very focused on software rather than system operation scenarios, thus realistic system testing scenarios are hardly achievable. 2) As the generated test suite is usually very large it is difficult to execute all test cases, especially in regard to the complete system. Test coverage is completely unknown if certain test cases are omitted.

## 3.3 Comparison of Related Approaches and Identification of the Research Gap

In Table 1, all previously presented approaches are compiled and rated regarding their fulfillment of the requirements imposed by the industry partners. As can be seen, none of the approaches is able to satisfy all requirements. Requirements-based test coverage approaches either cannot fulfill the required independence from formalized requirements or do not support test adequacy assessment (missing test cases or untested code does not become apparent). Code coverage-based approaches are mostly developed for the field of desktop computer software and not adjusted to the needs of software and industrial scenarios within production automation. While some approaches and tools were tested for their execution time overhead, explicit influence on the real time capability of production automation systems can only be speculated. One approach (Berger et al., 2012; Prähofer et al., 2011) presents a tracing method seemingly suitable for aPS, but does not take structured system testing or test adequacy assessment into account. Thus, a research gap is the lack of suitable test adequacy assessment support for aPS engineering that addresses the independence from formalized requirements while taking into account prevalent software standards, industrial system testing scenarios and real time constraints.

**Table 1: Overview of related approaches of system test adequacy assessment and rating of industrial requirements for aPS**

| | APS software (R1) | Real time capability (R2) | Inclusion of hardware behavior (R3) | Manipulation of hardware or process during testing (R4) | No formalized requirements needed (R5) | Test adequacy assessment – finding untested behavior (R6) |
|---|---|---|---|---|---|---|
| Available Requirements Management Tools (DOORS, Polarion) | Yes | Yes, no influence | Yes | Yes | Yes | Very limited |
| (Siegl et al., 2011) | Embedded SW | Yes, no influence (MIL/HIL) | No | No | No | Yes |
| (Kormann & Vogel-Heuser, 2011) | Yes | Minimal influence | Simulation | Yes (simulated) | Yes | Very limited |
| (Hametner, Hegny, & Zoitl, 2014) | Yes | Yes, no influence (SIL/MIL) | No | No | Yes | Very limited |
| (Rösch & Vogel-Heuser, 2017) | Yes | Minimal influence | Yes | Partially | No | Limited |
| (Kumar et al., 2013) | Yes | Yes, no influence (HIL) | Simulation | Yes (simulated) | No | Limited |
| (Sinha et al., 2016) | Yes | Yes, no influence (HIL) | Simulation | Yes (simulated) | No | Limited |
| (Wu et al., 2007) | Embedded SW | Minimal influence | No | No | Yes | Yes |
| (J Krause, Herrmann, & Diedrich, 2008) | Fieldbus communication | Yes, no influence (HIL) | No | No | No | Yes |
| (Jee et al., 2010, 2005) | Yes | Unknown for larger systems | No | No | Yes | Limited |
| (Simon et al., 2015) | Yes | Minimal influence | No | No | Yes | Limited |
| (Bürdek et al., 2015; Lochau et al., 2014) | Yes | Influence unsure | No | No | Yes | Limited |
| (Berger et al., 2012; Prähofer et al., 2011) | Yes | Minimal influence | Yes | Yes | Yes | None |
| Available Code Coverage Tools (Clover, Bullseye, PurifyPlus) | No, higher languages (Java, C++, .Net) | Unknown for aPS software | Yes | Yes | Yes | Yes |

## 4 Concept for Coverage Assessment during System Testing of aPS Using Execution Tracing

The following sections describe the developed approach in detail. As shown in Figure 1, the concept consists of multiple processes (pictured as oblong, dark grey items) to be performed in order to calculate and assess test coverage. Based on an original control software program, an instrumented version of the program is generated. In this process, a dependency model and a trace point database are created for later use in the coverage assessment process. Using a previously developed concept for guided semi-automatic system testing (Ulewicz & Vogel-Heuser, 2016) system test cases are manually derived from the system requirements. This process can also be supported by structured methods such as the failure mode and effect analysis (FMEA), which is commonly used in safety validation of aPS in the field of food, beverages and pharmaceutical products. The test cases are stored in a model format and are subsequently embedded within the PLC software project, which can be executed on the target hardware. During the guided semi-automatic system test execution, execution traces are generated for each test case along with a test report including the outcome of all performed test cases. Using this information as well as the previously generated dependency model and trace point information, coverage within the project can be calculated and visualized. This represents a support for the tester, especially inexperienced personnel, in assessing the adequacy of the performed test cases and the possibility to identify untested behavior, although the set of performed system tests was initially seen as potentially adequate. Even if the definition of test cases was performed thoroughly with methods such as the FMEA, critical implemented functionality might still be unintentionally omitted due to human error. The test set as well as other analysis artefacts, such as the FMEA, can be subsequently improved. In the following sections, each process will be presented in detail.

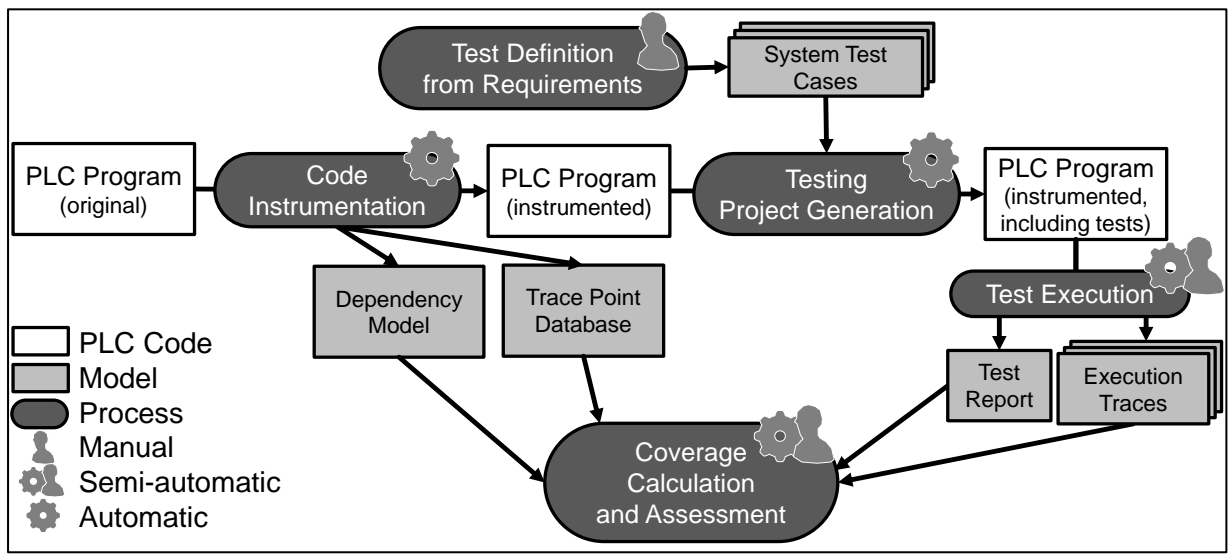

Figure 1: Concept overview of coverage assessment for aPS software: An original PLC program is automatically instrumented and extended by executable test cases. After semi-automatic system test execution, execution traces, a dependency model and information about generated trace points are used to automatically calculate test coverage, which supports testers in adequacy assessment.

### 4.1 Code Instrumentation

Instrumentation of the original control software program is required in order to record execution traces. For this, a suitable coverage metric, statement coverage, was chosen and a concept for code instrumentation and execution tracing using a dependency model of the control code was developed.

Choosing a Suitable Test Coverage Metric

As described in the section on related work, multiple different metrics have been developed in the field of computer science. Each metric has positive and negative aspects for the presented problem (identifying untested behavior), which were analyzed to choose a suitable metric with the given requirements of no influence on real time behavior and providing support in identifying untested system behavior.

Requirements based metrics are generally not suitable if detailed functional specifications are not available. A coverage cannot be calculated if the relation between detailed functional specifications cannot be made. In addition, unneeded functions, i.e. unneeded code, cannot be identified as these would not be specified even with detailed specifications. This type of coverage metric is therefore not suitable for the presented approach and was excluded from further consideration.

In contrast to this, code structure based metrics can be calculated without the need for additional detailed functional specifications. The identification of unimplemented yet required functions cannot be achieved with this type of metric, but this is not the goal of this approach. As stated in the section on related work, different metrics were developed in the field of computer science and checked for their suitability for assessing test suite adequacy, i.e. whether a test suit comprised of multiple test cases covers all relevant behavior in the system. Statement coverage was found to be very effective in detecting mutations, i.e. defects, in code (Gopinath et al., 2014). For identifying non-adequate test suites, i.e. test suites missing test cases, statement coverage does not seem to have any downsides compared to more detailed criteria (Gligoric et al., 2013). In addition, to record these detailed metrics, more detailed instrumentation is required: decisions need to be analyzed in more detail resulting in more runtime overhead and more memory is needed to store the information in case more complicated decisions are present. According to requirement R2, both available execution time and memory are critical and statement coverage is expected to require less of both in comparison to more complex metrics, such as condition/decision coverage. Industrial application was also expected to yield complex coverage results to be evaluated by the tester, which would be amplified by the even more detailed results from other metrics. Therefore, for this approach, statement coverage was more promising for the requirement of minimal influence on real time properties of the system and was subsequently chosen for the presented approach. Concepts for acquiring needed data were developed based on this choice. As stated in the outlook, an extension by or comparison to more detailed metrics, such as condition/decision coverage, is surely an interesting focus of future research.

## Code Instrumentation for Recording Test Coverage

For recording statement coverage, information about executed code statements during test execution, "execution traces" are needed. When regarding the system as a black box, this information is typically not available, i.e. not part of the interface of the system under test. In addition, a way of reliably recording this information, especially in a real time environment, poses certain problems: gathering this data must not miss executions of lines and must not change the system's real time behavior during recording. The two main problems to solve are therefore making the relevant information within the code observable and reliably and unobtrusively recording this information for later analysis.

This was achieved in the presented approach by analyzing the code and converting it to a system dependence graph ("Dependency Model") and subsequently instrumenting basic blocks (a set of statements without decisions, such as if-statements) within the code with "trace points": whenever a basic block is executed during execution a function call at the trace point is invoked to record and store this information. The code instrumentation was developed in such a way that minimal overhead regarding memory and calculation time was aimed for. The goal was to reduce the influence of real time properties of the system by the instrumentation in such a way that the instrumentation might remain in the final code, i.e. the code running on the final productive system during regular operation.

### Dependency Model

The dependency model used for identifying decisive points and basic blocks within the code is an extension of the dependency model definition presented in (Feldmann, Hauer, Ulewicz, & Vogel-Heuser, 2016). The original dependency model was designed to analyze programs for modularity and other maintainability properties by analyzing the control flow as well as the data flow within control programs. As this analysis did not consider the control flow within program organization units (POUs), the meta model, which is partially presented in Figure 2, was extended by suitable stereotypes to be able to represent these features in a generated model. The extended stereotypes are shown in light grey in the figure.

The dependency model is a directed graph consisting of nodes and edges. Nodes represent structural entities of a IEC 61131-3 project, whereas edges represent the dependencies between these entities (Feldmann et al., 2016). An edge connects two nodes, a source node and a target node, in one direction. The meta model is able to contain nodes from different hierarchies in the project, starting from the project itself, the defined tasks (threads) in the project, the POUs (functions, function blocks and functions) called by the tasks and other POUs and now even code elements such as actions (functions embedded in POUs) and basic blocks (code segments that do not contain decisions such as if-statements). For the programming language Sequential Function Chart (SFC), the "step" node was implemented, which can itself contain multiple actions per step. In the current version of the approach, actions with ST implementation and the type qualifiers P0 (single execution upon step deactivation), P1 (single execution upon step activation) and N (repeated execution while step being active) are supported (for all qualifiers, see (IEC, 2003)). The edges represent dependencies between the nodes, such as calls between POUs, write operations on variables and, in the extended model, also progressions between basic blocks (JumpsToEdge) and SFC-steps (SFCTransitionEdge). The "JumpsToEdge" is generated from control statements, such as if-statements, and additionally stores the condition as expressed in the if-statement or implicitly expressed in the else-statement. For example, if there is an if-statement, the "then"-part of the if- statement will be represented as an edge leading from the code before the "if" to the "then" part with the condition specified in the if-statement. If there is an "else" part of the statement, there will be an additional edge leading from the code before the if-statement to the "else" part with the inverted condition of the if-statement (see Figure 3 for a practical example). If a basic block calls another POU, there will be an edge leading from the first function block to the initial function block within the called POU without a condition. The same applies for the "SFCTransitionEdge": Transitions in SFC-charts are converted to transition edges, showing the connection between SFC steps as defined in the SFC chart.

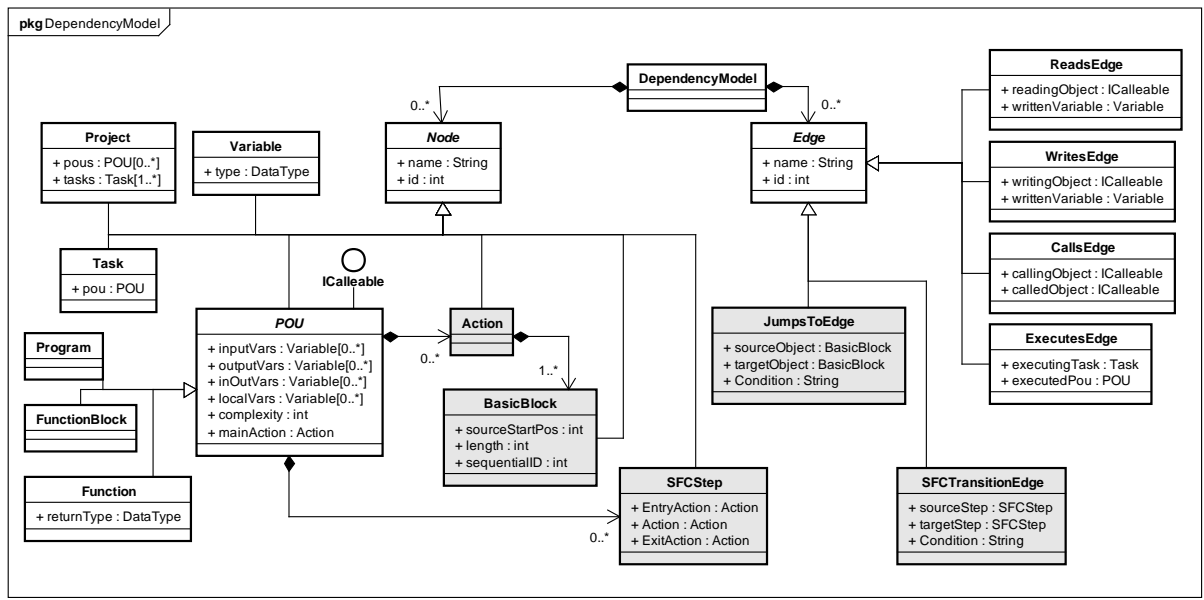

**Figure 2: Extended (grey shaded) dependency model for IEC 61131-3 based on** (Feldmann et al., 2016) **for code analysis**

*Creation of the Graphs*

Based on the meta model presented in the previous section, a dependency model is automatically generated from the source code of the PLC program. This is done by identifying all defined tasks as initial points for code exploration. The called POUs are identified from each initial point. The code specified within these POUs (its implementation body and, if available, its actions) is converted to an abstract syntax tree (AST), which is then walked through iteratively. During this walk-through, basic blocks and decisive points within the control flow are identified and saved as nodes and edges. In case a basic block calls another POU, this POU is equally walked through until the end of the code is reached. In this process, each basic block is sequentially numbered (*sequentialID : int*), to allow an easy correlation between basic blocks and traces data (see next section). The result is a call graph spanning the control software project's code which is relevant for the control flow, omitting POUs which will never be called. These POUs will already be identified by the compiler and in most cases directly marked within the IDE and are thus of no further relevance. The created graph is the basis for code instrumentation, explained in more detail in the next section.

*Inserting Trace Points and Creating a Trace Point Database*

The recording of data needed to infer execution traces is achieved by instrumenting the source code of the PLC project and saving information about inserted parts within a database. The instrumentation consists of inserting function calls in decisive points of the code and allocating memory for temporal storage of execution trace information. The trace information is realized as an array of Boolean variables for each decisive point in the code (*tpa : ARRAY[0..MAXTP] OF BOOL*). The array is reused for each test case by resetting each entry before each test case and saving the recorded information after each test case. For this, two functions and one function block were developed:

- Reset function *tp_reset()*: The reset function is called before each test case and is used to reset the complete trace array to ensure that all array items are set to their initial state (*false*).
- Record function *tpr(INT i)*: This function, which is called at each trace point, is given the identification number of the trace point after which the related array item in the trace array is set to "true".
- Saving function block *tp_save(BOOL xExecute, STRING szFilename)*: After each completed test case (failed or successful), this function will be called to save the information stored within the trace array into a common text file on the execution hardware. As this process might take several PLC scan cycles, the writing process needs to be completed (Output *xDone = TRUE*) before the next test case is initiated. The data saved is the id and value of each trace point (e.g. "1:true, 2:false, 3:false, …").

These POUs are inserted into the project alongside the trace array at the instrumentation phase of the control software project. Using the information collected in the dependency model, each basic block is instrumented with a trace point. Function calls of the tpr-Function are inserted into the code using the information about the location of the basic blocks (see Figure 3).

| Original code: | Instrumented code: |
|---|---|
| ```
1. IF in < 0 THEN
2.    out := -1;
3.    negative := TRUE;
4. ELSIF in = 0 THEN
5.    out := 0;
6.    negative := FALSE;
7. ELSE
8.    out := 1;
9.    negative := FALSE;
10.   END_IF
``` | ```
1. tpr(i:=42); IF in < 0 THEN
2.    tpr(i:=43); out := -1;
3.    negative := TRUE;
4. ELSIF in = 0 THEN
5.    tpr(i:=44); out := 0;
6.    negative := FALSE;
7. ELSE
8.    tpr(i:=45); out := 1;
9.    negative := FALSE;
10.   END_IF
``` |





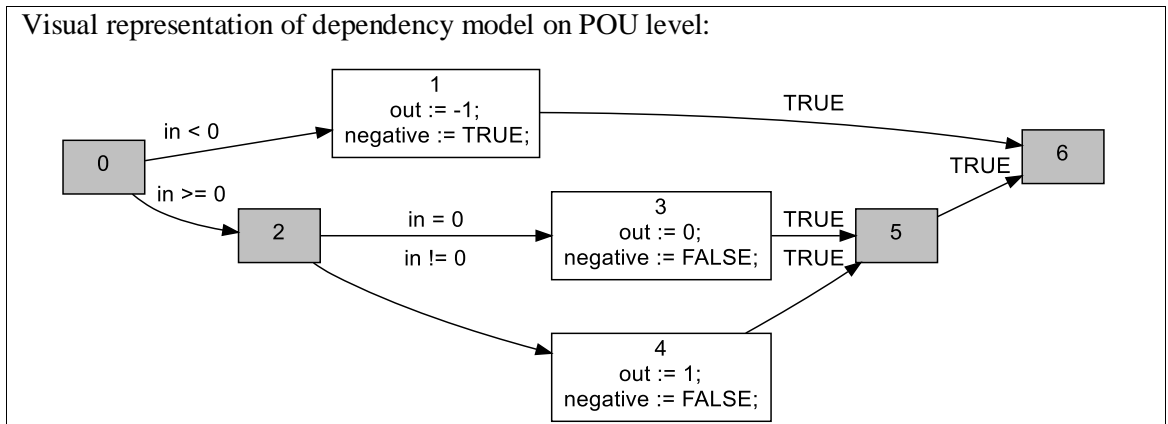

**Figure 3: Instrumentation example: The original code (upper left) is extended by function calls recording the execution of basic blocks (visual representation, bottom) resulting in instrumented code (upper right)**

Alongside the instrumentation of the code, a trace point database is created allowing the relation between the instrumented code, the dependency model and the execution trace information created during test execution. As shown in Figure 4, the database contains information about the related basic block and thus about the trace point location (sourceStartPos) of each inserted part within the code. In addition to this information, an entity named "visit" will be filled with information from the execution traces after the testing process is finished: each test case will create an execution trace stating if a trace point was "visited" or not. Thus, each trace point in the trace point database can be "visited" by each test case. This information is superimposed to identify which test cases were not visited by any test case ("WasVisited() = false") to identify untested parts of the source code.

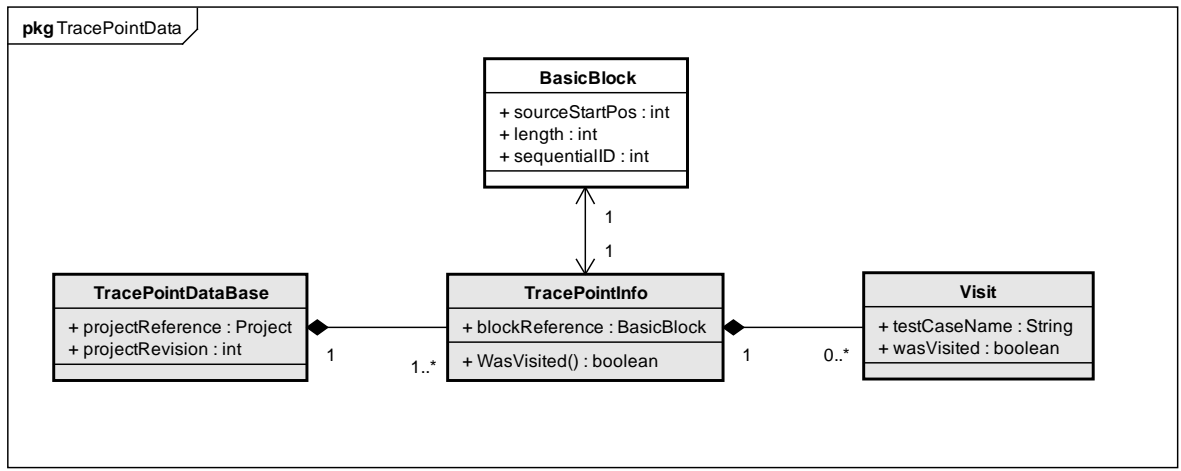

**Figure 4: Trace point database: Meta model to store information about instrumented points in the code and later relation to test case execution traces represented as "visits" or the respective trace point.**

## 4.2 Testing Project Generation and Test Execution

System testing for automated production systems can rarely be performed fully automatically: Many steps currently performed in system testing require a testing operator to stimulate the system and validate its behavior. During these manual processes, the relation between the performed actions and the conformance of the system behavior cannot be recorded and related automatically. Thus, neither the current approach of manual testing nor the automatic testing approach is suitable for application in the approach presented in this paper. For this reason, an approach for semi-automatic testing was introduced in previous work (Ulewicz & Vogel-Heuser, 2016). In this previous approach, the tester is embedded within a structured, guided system testing process. As many test steps as possible are performed automatically, while the remainder, where manual action of the tester is needed, are performed by the tester. The tasks the tester has to perform are displayed on an HMI, which is also used to give feedback to the testing system. In contrast to usual software testing, the parts of the system that can usually not





be stimulated or checked by the test bed can be included through intervention of the human operator. In addition, this approach enables the recording of execution traces during testing and a direct relation of these traces to the performed test cases. This approach is therefore a base for the coverage assessment.

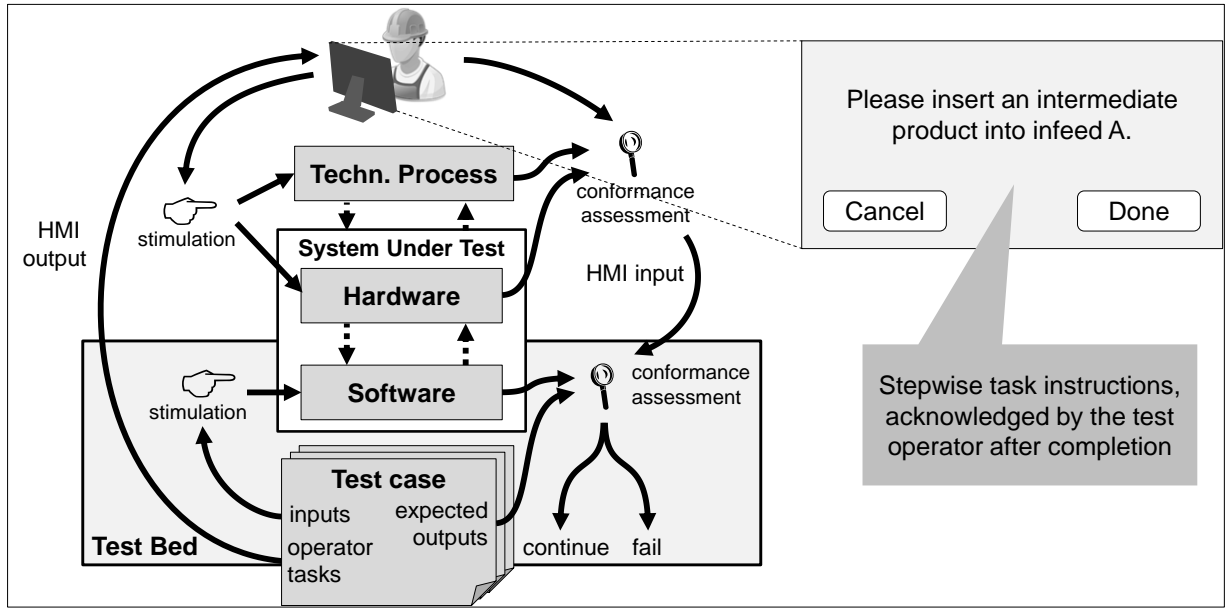

**Figure 5: Inclusion of the tester (engineer or technician) into the system testing process** (Ulewicz & Vogel-Heuser, 2016)**: By showing step-wise testing instructions on a display, the human can stimulate and check the system wherever the test bed cannot (hardware and technical process)**

The test cases, which are stored within the control software project, are defined in a model format (XML). Using the functionality provided by the CODESYS Test Manager (3S - Smart Software Solutions GmbH, 2016b), these test cases can be automatically converted into executable test cases. The tool can, among other things, upload the generated project onto any execution hardware, start the testing process, generate a test report and download any specified files from the execution hardware. This functionality was used in combination with the generation of additional tracing POUs and variables and the insertion of the HMI components for displaying test case information on a display and allowing user input during test execution, respectively.

The existing functionality was further extended for this approach by inserting calls into the test cases for invocation tracing function at the respective time. Additionally, the tracing functionality, which was previously used for tracing variable values during test execution, was extended by the execution tracing functionality, and the test script generation was adjusted to automatically download resulting execution trace files from the PLC to the development system.

## 4.3   Coverage Calculation and Assessment

As pointed out by Piwarowski (Piwowarski, Ohba, & Caruso, 1993) and Yang (Q. Yang et al., 2009), high coverage scores are difficult to achieve even regarding statement coverage. This may be due to unreachable code or complex conditions, among other reasons. This fact was also pointed out by the industry partners questioned in the initial requirements study: testing all behavior in every detail in an automated production system is not feasible. One reason why this is not possible in this particular field of industry is economics:  testing is resource intensive and performed under significant time pressure.





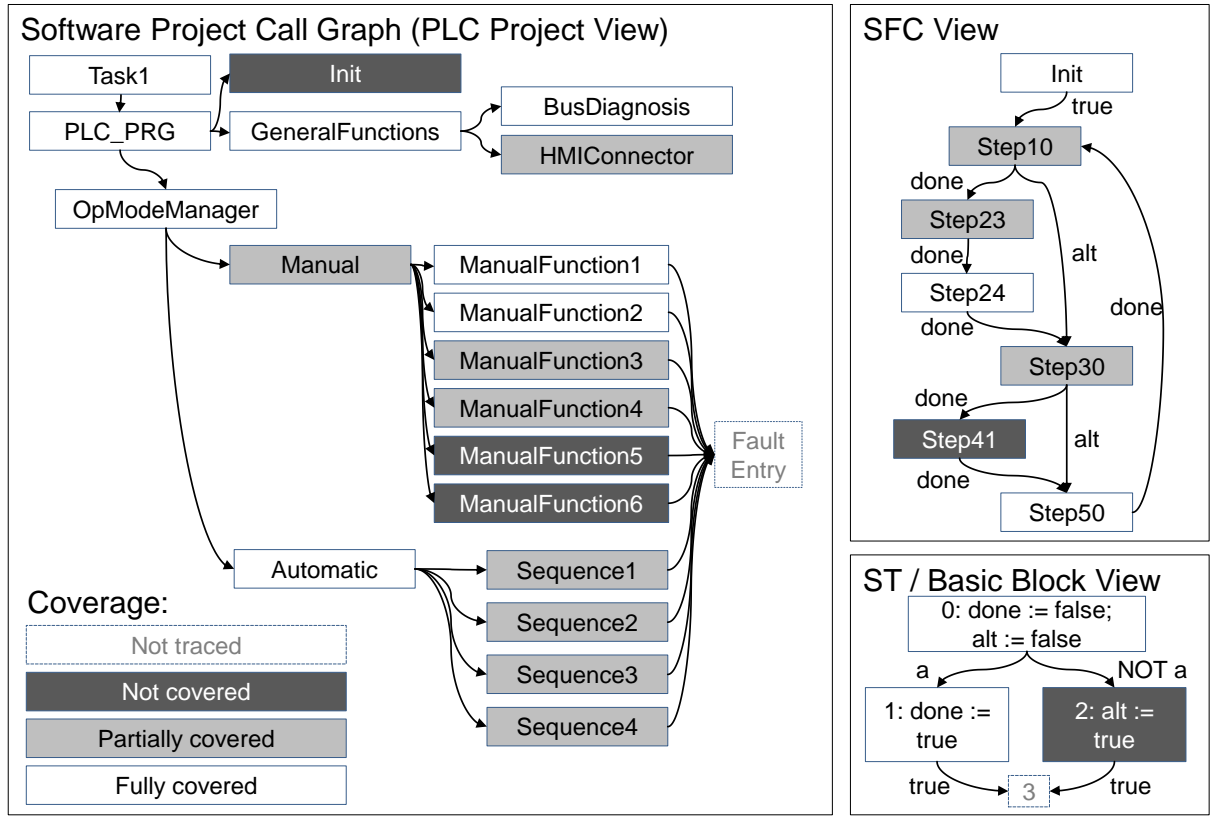

**Figure 6: Hierarchical coverage views as developed for the approach: A software project call graph (left) gives a quick overview of covered (light grey and white) and uncovered (dark grey) POUs. More detailed views can unveil uncovered parts of the code from SFC level (upper right) to ST level (lower right).**

As the goal of the presented approach is to identify untested behavior of the system, a quantitative measure as in a coverage percentage seemed unnecessary or unsuitable. Instead, a visual emphasis of untested code was chosen. This also allows the detailed investigation by the tester to evaluate whether the untested parts are indeed critical and therefore might require additional test cases. Inspired by a traffic light color scheme, untested parts are marked "red", i.e. need investigation, and partially tested parts are marked "yellow", i.e. potentially critical. A "green" marking was deliberately not used as parts of the system that were fully covered might still contain faults; many parts of the code are active during different system behaviors, which was deemed to be misleading.

For a quick assessment of the coverage of the system, different views were chosen aggregating the underlying coverage (see Figure 6). In a software project call graph, all executable POUs are depicted starting from the task calling the first POU. Each POU is marked "yellow" or "red" depending on the steps (in case of a POU programmed in SFC), actions or its basic blocks were only partially covered or not covered at all. More detailed views are presented by clicking on the respective POUs. In case of POUs programmed in SFC, the individual steps as well as their transitions are shown with a similar color coding. The level closest to the code is a view depicting individual basic blocks.

For future work, the authors think a direct implementation into the development environment's editors would result in the highest industrial acceptance of the approach as no new concepts would have to be learned. A mock-up of this idea is depicted in Figure 7.





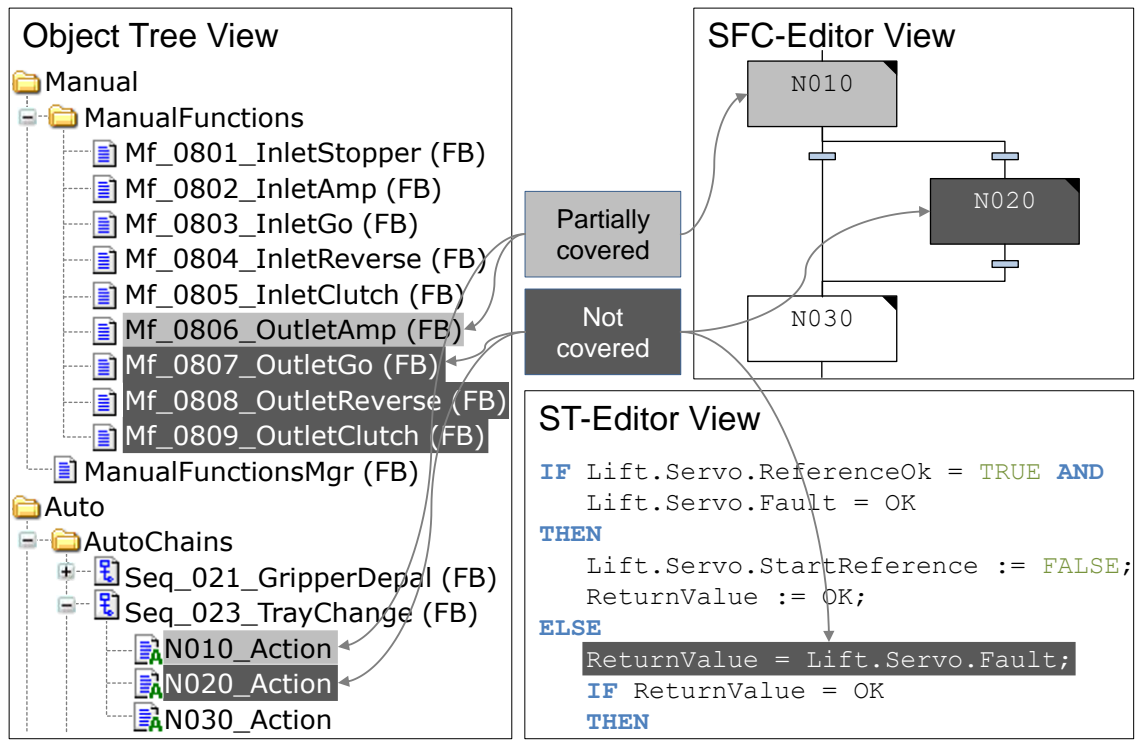

**Figure 7: Mock-up of editor integration of coverage visualization: Insufficiently covered items can be easily spotted, such as POUs in the object tree view (left), steps in the SFC-editor view (upper right) and basic blocks in the ST-editor view (lower right)**

By allowing the tester to quickly browse through the project to identify untested parts of the system, a quick ability to detect untested behavior of the code is expected. If a complete POU is marked as untested, the user can quickly look into the code and decide whether this block was previously tested or needs further investigation through additional tests. If an automatic step chain was only partially covered, the tester can identify the untested steps, which often correspond directly to behavior in the machine, and analyze the item for further investigation. This process can also be performed down to the basic block level where individual lines can be identified as untested, critical behavior or deliberately omitted.

# 5   Implementation

In order to be able to prove the applicability of the presented concept within the production automation domain, a prototypical tool for defining and executing tests and measuring and assessing test coverage was implemented. The tool was implemented as a plug-in for the widely used CODESYS V3.5 Integrated Development Environment (IDE) for automated production systems programmed in the IEC 61131-3 standard (3S - Smart Software Solutions GmbH, 2016a). Through the close integration with the IDE, information about the source code, its instrumentation and the automation of the test execution and coverage measurement was achieved. Using the capabilities of the IDE, dependencies and the abstract syntax tree could be easily extracted from the compile context, i.e. the object model used as an input for the compiler. The trace point database is saved as an XML-file for later use during the instrumentation.

The test definition and test project generation could be included from the previously developed semi-automatic system testing approach (Ulewicz & Vogel-Heuser, 2016), which was built upon the CODESYS Test Manager (3S - Smart Software Solutions GmbH, 2016b). Test generation was slightly extended by invoking function calls for resetting and saving the trace point data to execution trace files. These execution traces, saved between each test case, are stored directly on the embedded PC during test execution and are automatically transferred to the development system after finishing the testing process. The generated test script for the Test Manager was slightly adjusted for this process.





After test execution and download of execution traces, the developed plug-in automatically loads and analyzes the execution trace files coverage information and displays and browses this information visually.

## 6    Evaluation

Several experiments were designed with industrial experts and performed by the authors to evaluate the approach. The case study, the representative group of participants and the measured data were intentionally chosen as proposed by (Runeson, Höst, Rainer, & Regnell, 2012) to allow an evaluation of the initial requirements. During the experiments, execution time measurements were performed to assess the instrumentation's influence on runtime properties and thus its applicability for the production automation domain. In addition, the ability to perform coverage assessments was performed. The findings from the experiments were presented and discussed in a workshop with six experts active in the field of specially engineered automated production systems.

In the following sections, the system under test, the experiments and the results of the case study will be presented and subsequently discussed in relation to the requirements imposed on the approach in the first chapters of this work.

### 6.1    Case Study: The System under Test (SUT)

This system used for experimentation had been part of a real industrial factory automation system for depalletizing trays and passing the individual items on to the next station. Trays with parts are fed into the machine using conveyor belts. A lift system is used for locking the tray in position for picking and subsequent transport to a conveyor system transporting the empty tray out of the system. A 3-axis pick and place unit (PPU) is used to pick up individual pieces off the locked tray and place them into the next machine representing the next process step (this process step was not regarded in this work). A schematic view of the machine is depicted in Figure 8.

For interaction with the hardware (including 3 drives), 69 input variables and 26 output variables are defined (mostly Booleans). The control program was written by the company (not by any of the participants) and contained 119 program organization units and 372 actions adding up to about 15500 lines of code. The program includes two tasks, one for the control code, running at a PLC scan cycle time of 10ms, and one task for updating the visualization, running at 100ms. The programming languages used were IEC 61131-3 Structured Text (ST) and Sequential Function Chart (SFC). Thus, the size and complexity of the program represent a realistic application example. The program was initially written in the integrated development environment CODESYS V2 and ported to CODESYS V3.5 for this evaluation as the plug-in was developed for the newer version.

The system's control hardware is a Bosch Rexroth IndraControl VPP 21 embedded PC with a Pentium III 701MHz processor and 504MB of RAM. The embedded PC runs a CODESYS Control RTE V3.5.5.20 real time capable runtime. An Ethernet connection was used to connect the embedded PC to a development PC running the plug-in and uploading the test project to the embedded PC for real time capable execution.





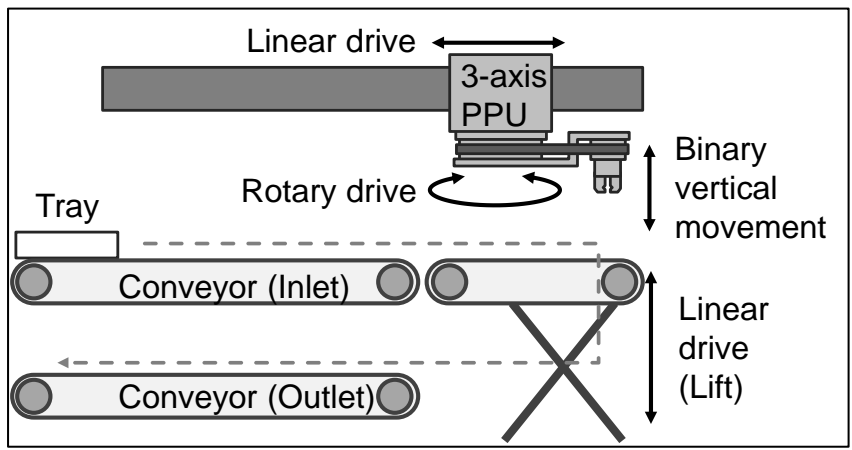

**Figure 8 Schematic view of the system under test (SUT)**

The development system used for generating the instrumented code, the test project and the coverage assessment was a consumer laptop with an Intel® Core™ i7 5600U CPU at 2.6GHz, 8GB RAM and running Microsoft Windows 10 64-bit. CODESYS V3.5 SP8 Patch 1 and CODESYS Test Manager Version 4.0.1.0. were used, including the developed plug-in.

## 6.2 Case Study: Description of Experiments and Measurement Results

The base for the experiments was a test suite of different system tests based on industrial system tests as currently manually performed. Using a generic test plan as defined for use in the cooperating company, 14 test cases were developed in cooperation with an industrial expert from this company and arranged into a test suite. The development of the test suite required about one hour, performing the test suite once required about 25 minutes including the generation of the test project. The test suite consisted of test cases testing individual manual functions, such as opening and closing the gripper using the HMI, automatic functions, such as regular operation with partially and completely filled trays, and special test cases, such as switching the operation mode during automatic operation.

Two experiments were performed: experiment I investigated the approach's feasibility and applicability of the coverage measurement qualitatively while experiment II was designed to investigate the properties of the overhead generated by the additional execution tracing.

Experiment I: Feasibility and Test Suite Coverage
The feasibility of the approach was investigated by measuring times for generation, the successful execution and tracing of the individual test cases and the coverage the test suite was able to achieve. The instrumentation of the project, resulting in 2261 trace points, took less than one second on each iteration. The complete test suite could be executed without problems, meaning that the test suite could be completed in its entirety without breaking real time restrictions (10ms for the main task) and with each execution trace being written completely onto the hard drive of the embedded hardware. Writing test traces was performed using an asynchronous writing function after test execution, which did not exceed 10 PLC scan cycles (the next test case was started after writing was finished not to influence the cycle time during test execution).

As coverage was never previously calculated or displayed, this was the second interesting property of the experiment. As expected, the test suite did cover most of the code but did not cover every detail although the test suite was designed according to the notion that most important behavior in the machine was included. Many manual functions were not covered, as no test was designed to specifically allow this, which was decided due to the similarity to the other test cases for manual functions. In real situations, these tests would have to have been specified. Some function blocks representing initialization functions were not covered as these were executed only once at program startup and thus not recorded during the actual test case execution. Some step chains were not covered as these represent behavior in case of emergency shut down. Most POUs regarding the behavior of the machine that were addressed by test cases were partially covered. In all cases, specific behavior of the system was not





included in the test case, mostly functionality related to fault detection. As an example, the behavior of the system in case of cycle time overrun was not investigated as no such errors occurred during test execution (see Figure 9). Another interesting finding was that unneeded code was detected: Due to time restrictions, several step chains were copied and modified resulting in complete branches of legacy SFC chains not being executed (see Figure 10).

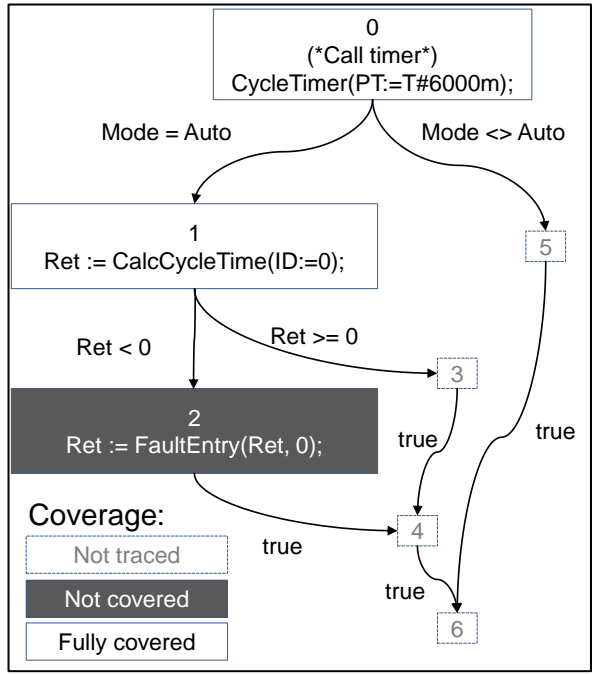

**Figure 9: Coverage assessment unveils untested fault handling routine: The code branch for documenting faulty cycle timing was never executed during the tests**

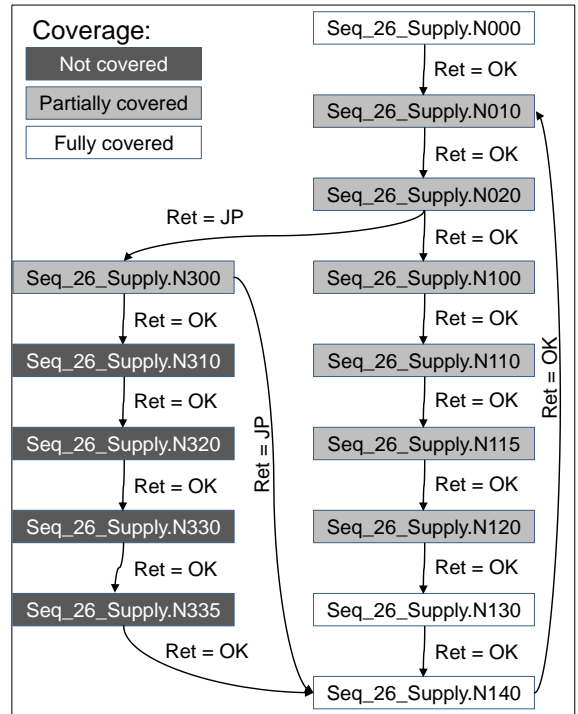

**Figure 10: Coverage assessment uncovers unneeded legacy code: A complete branch of the SFC code was never used and turned out to be obsolete upon closer inspection**

## Experiment II: Runtime and Memory Overhead by Execution Tracing

Two test cases were used in this experiment, which represented situations of different code being executed: a test case for a manual function of the gripper and a test case for automatic operation with a partially filled tray. Each test case was repeated five times, each in different configurations of the system to assess the overhead associated with each change:

- Original (no change): In this case, the test case was performed as it is currently being done – the test was performed fully manually without the semi-automatic guide of the HMI or any coverage tracing.
- Guided Test: In this configuration, the test was performed according to the guided semi-automatic testing approach as presented in (Ulewicz & Vogel-Heuser, 2016). In this case, no tracing was performed to investigate the overhead created by this concept alone.
- Coverage Tracing: In this configuration, the test was performed as proposed in this approach – the test was executed using the semi-automatic system testing approach with additional tracing of coverage. This last configuration was to separate the overhead created by the guided test approach from the overhead created solely by execution tracing.

Two individual measurements regarding execution time and memory size were performed for each test case, configuration and experiment repetition. The actual scan cycle time was measured during the test execution itself using the task monitor. This time includes reading inputs, executing the program and writing outputs and is relevant for the investigation on real time properties of the approach: if this time is below the desired PLC scan cycle time, real time properties can be held. For each configuration, the measurements for both test cases were averaged and the highest value was noted. Additionally, the size





of the compiled project divided into required memory for the compiled code and the global variables was noted. As this information did not change between repetitions, this measurement was only performed once for each configuration. The results of the measurements for the different configurations with the SUT are shown in Figure 11 for the execution time overhead and in Figure 12 for the overhead in required memory.

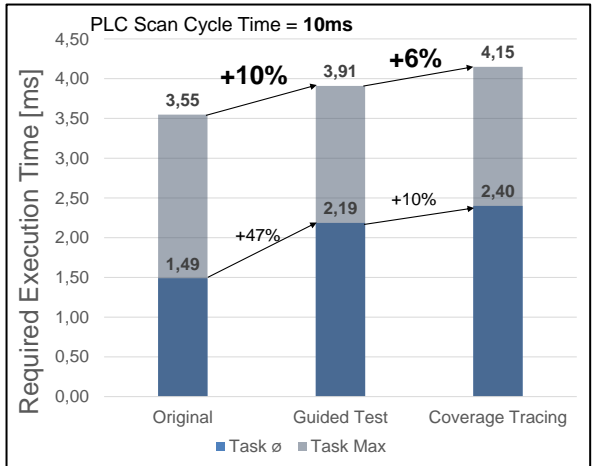
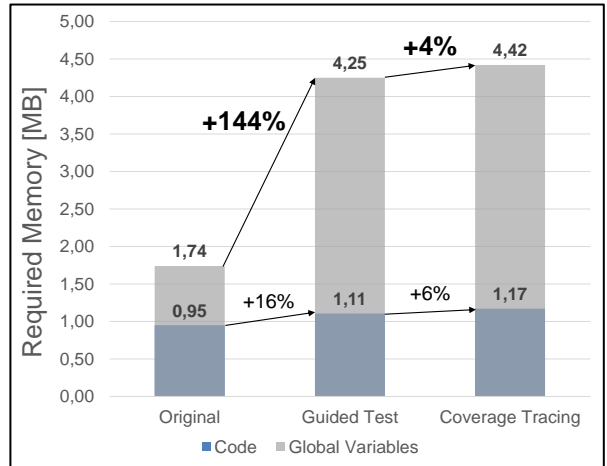

**Figure 11: Execution time overhead for SUT (Fig. 8)**  **Figure 12: Memory overhead for SUT (Fig. 8)**

The average scan cycle time increases significantly due to the guided testing concept (+47%), while the maximum time only increases moderately (10%). Although the increase in average execution time is unfortunate, the increase in maximum execution time is relevant for influence on the real time properties: real time properties can be held if the execution hardware and desired PLC cycle time allow for an increase of 10% in this case. The increase in average execution time is most likely due to an asynchronous, low priority writing operation, which was used to record specified I/O variables during execution.

The increase both in average and maximum scan cycle time is moderate regarding the addition of the execution tracing concept. The maximum execution time only increases by 6%, adding up to about a 16% increase in maximally needed scan cycle time due to the approach. Given the system under test (see Figure 8) for this evaluation with a required PLC scan cycle of 10ms, this increase would not influence the real time properties of the system.

The increase in required additional memory is quite low regarding the compiled code (+16% and +6%), yet the proportionate increase for globally needed variables for the guided test cases is very high (+144%). Although this increase is very high, the totally required memory is still very low (less than 5MB). The increase for the implementation of the execution tracing results in a slight increase of 4%.

## 6.3 Evaluation of Results with Industry Experts

The results of the measurements as well as the approach itself were discussed and evaluated in a group of six experts in the field of automated production systems. The group comprised employees active in the fields of commissioning, maintenance, aPS software engineering and group management (technical development) from the company engineering the machine used in the case study. The measurements in the previous section were presented to the group and subsequently discussed in terms of the requirements initially imposed on the approach. In addition, a questionnaire was filled out by each expert to quantify the results. Certainly, the group size does not allow a quantitative rating of the approach although qualitative conclusions were rendered a bit more precisely.

### Requirement 1 (R1): Support of Industrial Software Properties

As the system under test used in the case study was provided by the company as a representative example, the applicability of the support of industrial software properties was agreed upon by the





experts. The programming languages used in the case study are the only programming languages used by this company.

## Requirement 2 (R2): Real Time Capability and Memory Size

While the overhead in execution time did not represent a problem in the case study, the experts agreed that this fact could be problematic for machines with very short scan cycle times, e.g. highly automated mass production machines. In these cases, scan cycle times are kept as low as possible to increase production speed. Even slight increases in scan cycle time can result in noticeable and mostly unacceptable increases in production cycles (time needed for processing one product). This is due to SFC steps being executed for at least one scan cycle, with each increase in scan cycle time adding up for each step used in the production cycle.

To quantify the criticism mentioned by the experts in their questionnaire answers, it was estimated that the presented approach – in its current state – could only be applied to about 20%-33% of the machines produced by the company based on the increase in runtime overhead.

The overhead regarding memory was not seen as critical at all. Although the percentage increase seems large, current systems used by the company never ran into problems regarding memory. The experts estimated the approach to be applicable to about 90% of the machines produced by the company.

## Requirement 3 (R3): Inclusion of Hardware and Process Behavior

Using the guided, semi-automatic system testing approach, each test was performed including the real hardware and process properties, resulting in the fulfillment of the requirement.

## Requirement 4 (R4): Manipulation of Hardware and Process Behavior

The test cases were based on a test plan provided by the company and developed in cooperation with one of their experts, thus representing realistic test cases with realistic manipulation tasks. Therefore, the requirement for allowing the manipulation of hardware and process during test execution was approved. Still, it was noted that the resources needed to specify the test cases might not be reasonable for completely unique machines; reusing complex sub-modules several times might help to distribute the initial specification costs between multiple machines. According to the answers given in the questionnaire, it was estimated that the approach could be applied to 60% of the machines produced by this company.

## Requirement 5 (R5): No Need for Formalized Functional Requirements

In the case study, no additional formalized functional requirements for generating test cases regarding coverage criteria or to assess test coverage were needed, thus the requirement is fulfilled.

## Requirement 6 (R6): Support the Assessment of Test Adequacy

The coverage visualization was generally seen as very beneficial. While it was noted that marking the code as "tested" can be misinterpreted as "sufficiently tested", the positive aspects of the coverage assessment were agreed upon.

The experts evaluated the ability to quickly identify untested parts of the code as very beneficial. It had previously not been possible to get an overview of the executed parts of the code during testing, so test adequacy solely relied on the individual's estimation. Through easy identification of untested parts it is possible to (quickly) assess whether additional tests are needed and adjust the test suite accordingly. This results in the experts' opinion that the approach supports the assessment of test adequacy.

## Evaluation Conclusion

Summarizing the results of the case study and the expert evaluation, the approach was able to address all requirements while in some areas improvements could be made to extend the applicability in the field of production automation. Especially the overhead in runtime in the current version of the approach was seen as critical for some applications. On the whole, the experts fully agreed that the presented approach improves the ability to assess the code coverage compared to the current state (1.33 on a scale of 1 to 7, where 1 is fully agreed).





**Extrapolation of the Evaluation Results regarding Scan Cycle Time Overhead**

Subsequent to the presentation and discussion of the results of the evaluation, the reasons for the instrumentation overhead were investigated preliminarily. Theoretically, this limitation stems from two factors: remaining execution time in each PLC scan cycle and computational expense of the code regarding the currently chosen path through the control flow. With more control statements, more code instrumentation is required, requiring more execution time, as more inserted tracing functions need to be executed. This might lead to breaking real time requirements if not enough remaining execution time is available. Thus, the limit cannot be directly related to the code size, but might have to be calculated for each individual application using worst-case execution time analysis (Wilhelm et al., 2008), e.g. using static code analysis. Yet, code structure is not the only factor influencing execution properties, but also task scheduling, caching and scan cycle time jitter. In practice, the instrumented code could be treated as regular (more computationally expensive) code and enforcing a minimum remaining execution time, e.g. using a maximum of 80% of scan cycle time as a practical measure to avoid breaches of hard real time.

To gain a better understanding about the scalability of the approach, the data acquired during the case study was extrapolated for other scan cycle times and trace function invocations. With the given code, consisting of 119 POUs with an average cyclometric complexity of 9.47 and 372 actions with an average complexity of 2.13, 728 statements were executed on average during the execution of the test cases used in the evaluation. Using the average measured number of trace function invocations per scan cycle (395) and the average increase of scan cycle time attributed to the approach (0.26ms) with the given setup (701MHz, scan cycle of 10ms), Table 2 was created: This data represents an estimation of the average scan cycle time solely required for the invocation of trace function calls in regards to the amount of invocations and other scan cycle times. As expected, the percentage of required time rises with the number of trace function calls and shorter scan cycle times. Whether the remaining scan cycle time suffices for holding real time requirements (reading and writing input and output variables and executing the rest of the statements) strongly depends on the amount of statements invoked in a worst-case scenario (most computational intensive path through the code, including the trace function calls). This property is very specific for different aPS and their control software.

**Table 2: Percentage of PLC scan cycle time required for coverage tracing with the setup used in the case study**

| | PLC scan cycle time | | |
|---|---|---|---|
| **Trace function calls per scan cycle** | **10ms** | **5ms** | **1ms** |
| 10 | 0.05% | 0.11% | 0.54% |
| 50 | 0.27% | 0.54% | 2.72% |
| 100 | 0.54% | 1.09% | 5.44% |
| 200 | 1.09% | 2.18% | 10.89% |
| 300 | 1.63% | 3.27% | 16.33% |
| 400 | 2.18% | 4.35% | 21.77% |
| 1000 | 5.44% | 10.89% | 54.43% |

While this extrapolation gives an idea about the scalability properties of the approach, further investigation would be beneficial. Yet, many factors attribute to the real time capability of a system, as mentioned above.

## 6.4 Threads to Validity

The requirement analysis was performed with experts from three companies. The resulting requirements might not be applicable for some companies in the field of automated production systems. From the authors' experience and discussions with many experts from different companies from the production automation domain, the collected requirements are relevant for a large segment of this field of industry. The absence of formal specifications and simulations and the need for an improvement in system testing and its coverage assessment were mentioned to be of importance by the vast majority of industry partners.





The case study was performed only on a single machine. This small sample size does not permit making conclusions about scalability of the overhead or other properties that might prohibit the approach to be directly applied. In particular, aPS with significantly more control statements might yield an increased runtime overhead, due to more required tracing function calls. In particular for aPS controllers being operated near their full capacity (e.g. over 80% of available CPU time), this might lead to problems unless the PLC scan cycle time is increased.

The expert evaluation was performed with experts from a single company. The outcome of the questionnaire might differ if experts from other companies had been involved. To the best knowledge of the authors, the case study is comparable to a significant part of automated production systems and thus the evaluation can at the least provide a qualitative impression on the applicability of the approach.

The implementation of the approach was performed in a CODESYS-based environment with a real time capable embedded PC as execution hardware. Although the approach was developed for the programming standard IEC 61131-3, execution hardware and integrated development environments differ between the different vendors of automation hardware. For Siemens-based systems in particular, it is the authors' opinion that several problems might have to be overcome, such as creating the abstract syntax tree, saving data during runtime and automating the overall testing process. Yet, to the best knowledge of the authors, these problems should be possible to solve.

## 7 Conclusion

An approach for supporting the assessment of test adequacy for system testing in production automation was presented. For the first time valuable support in assessing and increasing testing quality in fully integrated industrial production automation systems is presented while taking strict industrial requirements and quality assurance scenarios into account. Notably, acceptable influence on real time constraints could be achieved and no formalized requirements specifications or other elaborate engineering artefacts are needed. In addition, the performed test cases include the system's real hardware and technical process behavior and therefore allow realistic system tests through the partial inclusion of human testers within the system testing process.

The approach focuses on the identification of untested code and its relation to untested behavior of the system. The requirements relevant for this approach were gathered in collaboration with experts from the industry. Based on these requirements, an approach was developed that uses code instrumentation for tracing code execution during runtime while test cases are performed. The information is subsequently gathered and used for calculating and visualizing coverage of the performed test suite. Through this process, it is possible to quickly identify untested parts of the code and decide accordingly whether additional tests are needed.

The approach was implemented in an industrial case study, assessing its applicability with a group of experienced experts active in the fields of aPS commissioning, maintenance and software engineering. The approach is not applicable on all machines in its current state, mostly due to the (currently not optimized) runtime overhead derived from the code instrumentation.

Despite the discussion about the runtime overhead, it was agreed by the experts that the approach allows for the first time support of the tester in assessing test coverage and identifying untested behavior of the system and is applicable for a significant proportion of machines produced by this company.

As the experts' requirements and machines are similar to a substantial part of the authors' industry partners, the approach represents a significant improvement in supporting the testing process within the production automation domain.

## 8 Outlook

Multiple threads of further research were identified given the results of the evaluation and experience gained during the experiments.





As for the main criticism of the runtime overhead, multiple improvements could be made and will be pursued in future work. For example, the amount of trace points could be reduced by assuming deterministic execution of code within one task, removing tracing else-blocks in if-else-constructions: If the previous block was executed and the basic block within the if-statement was executed, tracing of the else-block can be omitted as these blocks mutually exclude each other. Another example could be optimizing the I/O-tracing algorithms used in the guided testing approach which was a source of a large increase in average execution time.

Statement coverage as the coverage measure was mostly sufficient in the case study, yet in certain cases, a closer investigation of the decisions in the if-statements would have been of interest. For this reason it might be interesting to look into a comparison of this metric with a version of condition/decision coverage in terms of level of detail and additionally required performance based on the more complicated instrumentation. In addition, for measuring coverage of functions, which are not predominantly complex because of decisions, e.g. mathematical or closed loop controller functions, other coverage measures should be investigated for more meaningful coverage results.

So far only the IEC 61131-3 languages Sequential Function Chart (SFC) and Structured Text (ST) are supported. In future work a coverage measurement and visualization of the visual programming languages Function Block Diagram (FBD), Ladder Diagram (LD) and Continuous Function Chart (CFC), and further consideration of all types and implementations of SFC actions, e.g. timed actions or action implementation in programming languages other than ST, might be of interest.

## Acknowledgement

The authors would like to express their gratitude to the companies 3S – Smart Software Solutions GmbH, Bosch Rexroth AG and Robert Bosch GmbH for helpful discussions and their support in optimizing and evaluating the presented approaches.

This work was supported by the Bavarian Ministry of Economic Affairs and Media, Energy and Technology within the research program "Informations- und Kommunikationstechnik in Bayern" [grant number IUK413].